\documentclass{iau}
\usepackage{graphicx} 

\title[IAUS291.~~Surrounded by spiders] %% short title %%
{Surrounded by spiders! New black widows and redbacks in the Galactic field} %% full title %%

\author[M. S. E. Roberts]  %% short author list %%
{Mallory S.E. Roberts}
\affiliation{Eureka Scientific, 2452 Delmer Street, Suite 100, Oakland, CA 94602-3017, USA \\[\affilskip]  Department of Physics, Ithaca College, Ithaca, NY 14850, USA \\[\affilskip] email: {\tt malloryr@gmail.com} \\}
\pubyear{2012}
\volume{291}  %% insert here IAU Symposium No.
\jname{\mbox{Neutron Stars and Pulsars: Challenges and Opportunities after 80 years}}
\editors{J. van Leeuwen, ed.} 
\begin{document}

\maketitle

%% -- Abstract ----------------------------------
\begin{abstract}

Over the last few years, the number of known eclipsing radio millisecond pulsar systems in the Galactic field has dramatically increased, with many being associated with Fermi gamma-ray sources. All are in tight binaries (orbital period $< 24$ hr) with many being classical ``black widows" which have very low mass companions (companion mass $M_c << 0.1 M_{\odot}$) but some are ``redbacks" with low mass ($M_c \sim 0.2-0.4 M_{\odot}$) companions which are probably non-degenerate. These latter are systems where the mass transfer process may have only temporarily halted, and so are transitional systems between low mass X-ray binaries and ordinary binary millisecond pulsars. Here we review the new discoveries and their multi-wavelength properties, and briefly discuss models of shock emission, mass determinations, and evolutionary scenarios.
\end{abstract}

% add below any authors, subjects and objects for indexing 
%   add more lines if necessary
%   but leave all lines commented out
%\index[author]{Roberts, M.S.E.|textbf}
%\index[author]{LastName2, Initials|textbf}
%\index[subject]{Keyword1}
%\index[subject]{Keyword2}
%\index[object]{PSR J2129-0429}
%\index[object]{Object2}

\firstsection % if your document starts with a section,
\section{Introduction}

In 1962, Giacconi and collaborators discovered the first low mass X-ray binary (LMXB) Sco X-1,  thought to be a neutron star in an 18.9~hr orbit accreting from a $\sim 0.4 M_{\odot}$ companion (Steeghs \& Casares 2002).
Shortly after the discovery of the first millisecond radio pulsar (MSP) PSR B1937+21, Alpar et al. (1982) proposed that radio MSPs were the end product of the accretion process we are observing in LMXBs. 
Of the roughly 2000 radio pulsars known today, about 10\% are MSPs, old neutron stars which have been
spun-up, or ``recycled", through accretion of
material from a companion. Many details of this recycling process remain unknown, but it is clear that
most  fast-spinning ($P < 8$~ms) MSPs  have degenerate white dwarf
companions with masses between 0.1 and 0.4~$M_\odot$, and very tightly follow the expected correlation between companion mass and orbital period derived from binary evolution models (see Figure 1).
However, some (up to 20\%) of MSPs are isolated. The process through which these MSPs were formed is unclear.
Did they somehow lose their companions or were they born through a different formation channel? 
One idea is that the companions of isolated MSPs were ablated away by energetic particles and/or $\gamma$-rays produced by the pulsar wind. This idea  was  inspired by  the discovery of the original ``Black Widow" pulsar, B1957+20 (Fruchter, Stinebring \& Taylor 1988), an eclipsing 1.6~ms pulsar in a 9.1-hr orbit around a very low mass ($M_C\sim 0.02 M_{\odot}$) companion. 

%% CUP work flow only accepts EPS -- not PDF, JPG, etc.
\begin{figure}[h]
\begin{center}
\includegraphics[height=3.4in,angle=90]{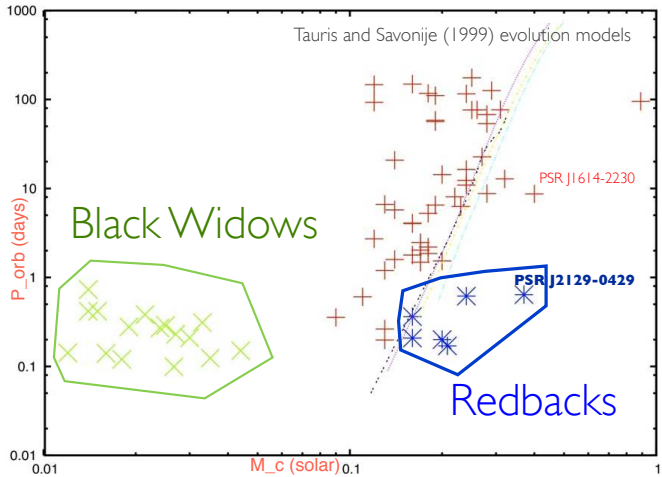} 
\caption{Minimum companion mass vs. orbital period of fast ($P < 8$~ms) binary MSPs in the Galactic field, showing the positions of the Redbacks and Black Widows discussed in this proceedings. The lines are from various binary evolution models considered by Tauris \& Savonije (1999)  which result in a Helium white dwarf companion. Since we plot minimum companion masses, we expect systems which evolved according to standard evolutionary scenarios  be just to the left of these lines. Note the one clear, non-spider exception at multi-day periods is PSR J1614$-$2230, which has a CO white dwarf companion and is a major challenge to current models (eg. Tauris, Langer \& Kramer 2011). Non-spider MSPs (plus signs) are from the ATNF pulsar catalog {\tt http://www.atnf.csiro.au/research/pulsar/psrcat} }
\label{fig1}
\end{center}
\end{figure}

The accretion formation scenario has been strongly supported by observations of fast X-ray millisecond pulsations in transient LMXBs and Type I X-ray bursts (see ``Compact Stellar X-ray Sources" eds. Lewin and van der Klis for reviews).
The ``missing link" of the MSP formation scenario, PSR J1023+0038,  was discovered in 
a Green Bank Telescope drift scan pulsar survey (\cite[Archibald et
  al.\ 2009]{asr+09}).  
This 1.69~ms radio pulsar is in a 4.8~hr orbit
around an $\sim 0.2M_{\odot}$ non-degenerate companion, and exhibits regular radio eclipses around superior conjunction. 
Optical spectra taken in 2001 of the system, before radio pulsations were discovered,  showed emission lines indicating the presence of an accretion disk but observations in 2004 showed only absorption lines suggesting the accretion disk 
had vanished (\cite[Wang et al.\ 2009]{wat+09}). 
X-ray studies with XMM-Newton (Archibald et al. 2010) and Chandra (Bogdanov et al.\ 2011) show significant orbital modulation in addition to pulsations, presumably arising from an intrabinary shock. 
The eclipse depth and duration imply that the shock is localized near or at the surface of the companion.
The energetics favor a magnetically dominated pulsar wind that is focused into the orbital plane, requiring close alignment of the pulsar spin and orbital angular momentum axes as expected. The X-ray spectrum consists of a dominant non-thermal component from the shock and at least one thermal component, likely originating from the heated pulsar polar caps.  This source has become the prototype of the ``redback" systems, named after the Australian cousin to the North American black widow.  Although having similar energetics, orbital periods, and eclipses, redbacks tend to have companions whose masses are higher than the expected white dwarf endpoint for their orbital period, rather than much lower as for the black widow like pulsars (see Figure 1). 

For both PSR B1957+20 and PSR J1023+0038, orbital modulation of the illuminated optical companions show that they are nearly filling their Roche lobes (\cite[van Kerkwijk et
al. 2011, Deller et al. 2012]{vbk11,dab+12}). Optical studies have been used to estimate the inclination angles and masses of the neutron stars, which are well above the canonical $1.4 M_{\odot}$. 

In principle, these systems should be excellent for studying the acceleration,  composition, and shock dynamics of the highly relativistic winds coming from energetic pulsars (\cite[Arons \& Tavani 1993]{at93}). In particular, in terms of light cylinder radii, the intrabinary shocks in these systems are tens of thousands of times closer than the termination shock of young pulsar wind nebulae like the Crab. This means that we can probe  the pulsar wind in regions where the magnetization parameter of the ultra-relativistic wind $\sigma$ may still be relatively high. 

%%Takata, Cheng, \& Taam (2012) recently estimated the expected X-ray and $\gamma$-ray emission from these intrabinary shocks, and find that the X-ray luminosity should be strongly correlated with the magnetization parameter. However, the effects of synchrotron beaming and bulk flow doppler boosting need to be taken into account as well, meaning that orbitally resolved spectral studies are required if we want to use X-ray emission as a probe of the pulsar wind. We also need to disentangle thermal emission from the neutron star surface and magnetospheric emission from the shock emission. Of these three potential sources of X-ray emission, only the intrabinary shock is expected to be orbitally modulated, unless the neutron star is actually eclipsed by the companion (note the extended radio eclipses are primarily due to scattering by a very modest amount of intrabinary material). Therefore, a high enough signal to noise between the brighter and fainter orbital phases in a number of systems is crucial for quantitative studies of how such emission depends on spin-down energy and binary separation.

Deep pulse searches of globular clusters over the years have yielded 18 binary 
pulsars with very low, black widow like, companion masses. 
They also yielded 12 short orbit eclipsing systems with more ordinary companion masses often showing optical evidence of a non-degenerate companion, i.e. redback like systems (see P. Freire's webpage {\tt http://www.naic.edu/$\sim$pfreire/GCpsr.html} for a complete list of globular cluster MSPs). 
Unfortunately, studies of the energetics, X-ray properties, and optical companions of globular cluster pulsars are difficult due to the crowded fields, the on average large distance, and the gravitational well of the cluster often making the intrinsic spin-down rate impossible to infer. Nearby systems in the Galactic 
field are therefore desirable for detailed studies.

%%%%%%%%%%%%%%%%%%%%%%%%%%%%%%%%%%%%%%%%%%%%%%%%%%%%%%%%%%%%%%%%%%%%%%%%%
%%
%%   use this format to include a LaTeX table  into your paper
%%
\begin{table}[h!]
\begin{tabular}{lcccccc}
\hline
Pulsar$^1$ &  $P_s$ & $\dot E/10^{34}$ $^2$ & $d_{NE2001}$ &
$P_B$ & $M_c$ $^3$ & ref. \\ 
 & (ms) & (erg/s) & (kpc) & (hr) & (solar) \\ \hline
\multicolumn{6}{c}{Old Black Widows} \\
\hline
B1957+20 F &  1.61  &  11  & 2.5 & 9.2 & 0.021 & \cite{fbb+90}\\
J0610$-$2100 &  3.86 & 0.23 & 3.5 & 6.9 & 0.025 & \cite{bjd+06} \\ 
J2051$-$0827 & 4.51 & 0.53 & 1.0 & 2.4 & 0.027 & \cite{sbl+96}\\
\hline
\multicolumn{6}{c}{New Black Widows} \\
\hline
J2241$-$5236 F & 2.19 & 2.5 & 0.5 & 3.4 & 0.012 & \cite{kjr+11} \\
J2214+3000 F & 3.12 & 1.9 & 3.6 & 10.0 & 0.014 & \cite{rrc+11} \\
J1745+1017 F & 2.65 & 1.3 & 1.3 & 17.5 & 0.014 & \cite{b+12}\\\
J2234+0944 F & 3.63 & 1.6 & 1.0 & 10 & 0.015 & \cite{k+12} \\
J0023+0923 F & 3.05 & 1.6 & 0.7 & 3.3 & 0.016 & \cite{h+11} \\
J1544+4937 F & 2.16 & 1.2 & 1.2 & 2.8 & 0.018 & \cite{bg+12} \\
J1446$-$4701 F & 2.19 & 3.8 & 1.5 & 6.7 & 0.019 & \cite{kjb+12} \\
J1301+0833 F & 1.84 & 6.8 & 0.7 & 6.5 & 0.024 & \cite{rap+12} \\
J1124$-$3653 F & 2.41 & 1.6 & 1.7 & 5.4 & 0.027 & \cite{h+11} \\
J2256$-$1024 F & 2.29 & 5.2 & 0.6 & 5.1 & 0.034 & \cite{blm+11} \\
J2047+10 F & 4.29 & 1.0 & 2.0 & 3.0 & 0.035 & \cite{rap+12} \\
J1731$-$1847 & 2.3 & ?? & 2.5 & 7.5 & 0.04 & \cite{b+11} \\
J1810+1744 F & 1.66 & 3.9 & 2.0 & 3.6 & 0.044 & \cite{h+11} \\
\hline
\multicolumn{6}{c}{New Redbacks} \\
\hline
J1628$-$32 F & 3.21 & 1.8 & 1.2 & 5.0 & 0.16 & \cite{rap+12}\\
J1816+4510 F & 3.19 & 5.2 & 2.4 & 8.7 & 0.16 & \cite{ksr+12}\\
J1023+0038 F & 1.69 &$\sim 5-10$ & 0.6 & 4.8 & 0.2 & \cite{asr+09}\\
J2215+5135 F & 2.61 & 6.2 & 3.0 & 4.2 & 0.22 & \cite{h+11}\\
J1723$-$28 & 1.86 & ?? & 0.75 & 14.8 & 0.24 & \cite{clm+10} \\
J2129$-$0429 F & 7.61 & 3.9 & 0.9 & 15.2 & 0.37 & \cite{h+11}\\
\hline
\end{tabular}
\caption{Black Widows and Redbacks in the Galactic Field}
\label{tab:bws}
$^1$ an F indicates a Fermi source. $^2$ assuming $1.4M_{\odot}$ and 10~km radius. $^3$ assuming $1.4M_{\odot}$ pulsar and $i=90^{\circ}$.
\end{table}
%%%%%%%%%%%%%%%%%%%%%%%%%%%%%%%%%%%%%%%%%%%%%%%%%%%%%%%%%%%%%%%%%%%%%%%%%%%

However, for the 20 years between the discovery of PSR B1957+20 and PSR J1023+0038, eclipsing MSPs remained rare outside of globular clusters. Only two black widow like systems were discovered during this time period, both with relatively low spin-down energies. 
Radio surveys sensitive to fast pulsars in tight binaries have recently
greatly increased the number of known black
widows and redbacks in the Galactic field (see Table 1). Surveying high-latitude Fermi $\gamma$-ray sources has proven to be the most productive means of discovering new MSPs, with 43 discovered so far (Ray et al. 2012). A surprising fraction of these new MSPs have turned out to be eclipsing sources in compact ($< 1$~day)
orbits. All told, there are more than 12 new black widows and 6 new 
redbacks that have been discovered in the Galactic field over the last 3 years. 

Timing of these systems is complicated by a tendency for their pulse profiles to vary as a function of frequency. For example, the redback PSR J2215+5135 
discovered in a 350~MHz Green Bank Telescope survey of Fermi sources (\cite[Hessels et al. 2011]{h+11}) shows a double pulse profile at 350~MHz, but by 2 GHz is predominantly single peaked, which peak is neither of the peaks seen at 350~MHz. Eclipse fractions also vary significantly as a function of frequency. Most of the systems show some amount of change in orbital period, presumably from the  mass quadrapole moment of the companion. Long term monitoring of these orbital variations may serve as a unique probe of the stellar structure of their companions.  
Optical studies of the companions show they are often close to filling their Roche lobes, and so can be significantly non-spheroidal (\cite[Breton et al. 2012]{bvr+12}). 

An unfortunate aspect of these changes in orbital period is that it is often problematic to extrapolate  timing solutions back to the beginning of the $Fermi$ mission to allow the most sensitive $\gamma$-ray pulse studies. Despite this, once a lengthy enough timing solution is obtained, folding the $Fermi$ $\gamma$-ray data has usually resulted in a pulsed detection above 100~MeV. 

X-ray studies of these new systems show that they tend to have an orbitally modulated, non-thermal component to their X-ray emission (\cite[Gentile et al. 2012]{gmr+12}). This is presumably due to the intrabinary shock. Current studies are statistically limited, however, and deeper X-ray observations are needed for detailed studies. So far, there has been no strong evidence of orbital modulation of the $\gamma$-ray emission detected by $Fermi$, and only upper limits have been obtained from TeV observations.

Optical studies have shown that both types of systems show evidence of bloated 
companions, with Roche lobe filling factors $\sim 0.4-1.0$ (\cite[Breton et al. 2012]{bvr+12}). By combining light
curve modelling and radial velocity measurements of photospheric
lines, the geometry of the systems and strong mass constraints on the
components can be obtained. The distinctive optical light curves can
be used to identify systems whose radio pulsations cannot be detected,
either due to radio beaming or having a 100\% eclipse
fraction. Indeed, there have already been a few such systems
discovered (\cite[Kong et al. 2012, Romani \& Shaw 2011]{khc+12,rs11}).

Neither black widows nor redbacks seem to be the end products of
standard binary evolution models. However, at least for redbacks, it
seems clear that the companions are not yet degenerate, and so may be
systems where the recycling process is temporarily halted, but will
resume in the near future. The position of redbacks on the
companion mass--orbital period diagram (Figure 1) is intriguingly
consistent with model evolutionary tracks of systems whose endpoints
are ultra-compact X-ray binaries (\cite[Podsiadlowski et al. 2002]{prp02}). Black Widows, however, seem to occupy  positions on the diagram which no evolutionary tracks pass through. 

\begin{figure}[h]
\begin{center}
\includegraphics[height=3.4in,angle=90]{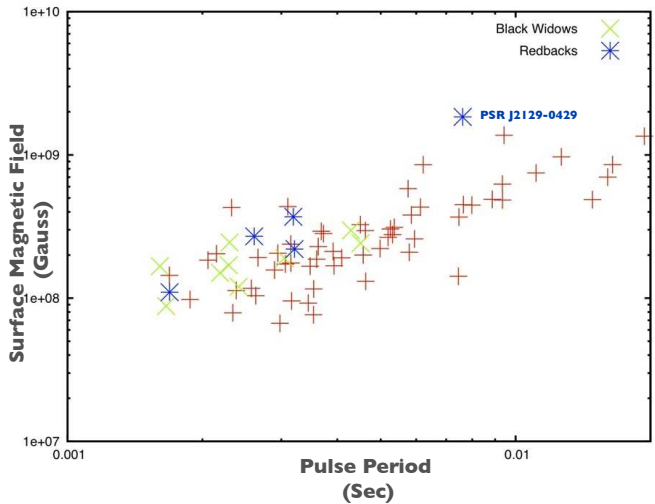}
\caption{Spin period vs. inferred surface magnetic field for fast MSPs in the Galactic field. In the case of no magnetic field decay
after accretion stops, pulsars near maximum spin-up should be to the left of average for a given magnetic field. Note that not all pulsars in this diagram have had their spin-downs corrected for their motion in the Galaxy, which means that their magnetic fields would be overestimated. Also, variations in moments of inertia (i.e. different pulsar masses) and magnetic inclination angles would also affect the inferred magnetic field in the standard dipole approximation.}
\label{fig2}
\end{center}
\end{figure}

The suppression and decay of the magnetic field from $B\sim 10^{12}$~G down to $\sim 10^8$~G during the accretion process is still not well understood, but must happen for pulsars to be spun-up to periods shorter than a few milliseconds (\cite[Cumming 2005]{c05}).  There is, however, an empirical relationship between the spin period of a MSP and its magnetic field inferred from its spin-down. For MSPs with white dwarf companions, billions of years may have passed since the recycling process ended, and significant spin down will typically have occurred. If redbacks are systems in which 
the recycling process is only temporarily interrupted, then we might expect them to be very nearly maximally spun-up for their magnetic field. In Figure 2, we show this empirical relationship, and indeed, redbacks tend to be among the fastest spinning of MSPs for a given magnetic field. Note that the magnetic field inferred from spin-down assumes pure dipole radiation, the canonical moment of inertia, and would in principle would also depend on magnetic inclination angle. 

One of the new redback systems is particularly noteworthy as having both the largest minimum companion mass and largest inferred magnetic field. PSR J2129$-$0429 was also discovered in the 350~MHz Green Bank Telescope survey of $Fermi$ sources. It shows extensive radio eclipses and has a bright UV counterpart seen with the Swift UVOT. The Swift X-ray data suggest strong orbital modulation, and that it may have the brightest X-ray shock emission of any of the known eclipsing binaries.
Its orbit and companion mass are actually quite similar to that of Sco X-1. If nearly filling its Roche lobe, the companion would only be $\sim 8000$ light cylinder radii away from the pulsar. Further studies of this system may give us a unique window into the earlier stages of the recycling process. 

The dramatic increase in the number of systems with intrabinary pulsar wind shocks may herald a new era of pulsar wind studies. The recent evidence of dramatic GeV-TeV flaring in the high mass radio pulsar binary PSR B1259$-$63 demonstrates the potentially strong geometrical dependence of high energy emission from pulsar wind shocks (\cite[Abdo
et al. 2011]{aaa+11}). While the binary separation and nature of the companion is very different between PSR B1259$-$63 and the eclipsing MSPs, the spin-down fluxes are within an order of magnitude, and the density of photons available for Compton up-scattering  from the companion at the shock is actually fairly similar. Therefore, they may prove to be a new class of TeV sources, either to the current generation of instruments or to CTA. 

The accelerated rate of discovery of new eclipsing systems is likely to continue for several more years, as more results come in from ongoing large scale pulsar surveys and more $Fermi$ sources are discovered and searched. Radio, optical, X-ray, and $\gamma$-ray studies all hold much potential for discovery. Well constrained pulsar masses, insights into the recycling process, unique tests of particle particle acceleration models, and even greater understanding of the structure of evolved low-mass stars may all result from current work on these new systems.

\end{document}